Title: Adiabatic theory of off-center vibronic polarons: Local dynamics and itinerant features along a linear chain
Authors: M. Georgiev and M. Borissov[†] (Institute of Solid State Physics, Bulgarian Academy of Sciences, Sofia, Bulgaria) ([†] deceased)
Comments: 14 pages pdf format
Subj-class: cond-mat

The vibronic mixing by an odd-parity vibrational mode of two opposite-parity nearly-degenerate electronic states at a molecular site may drive the small polaron off-center. An associated electric dipole may occur due to broken inversion symmetry which will exert an effect on the dielectric, optical and transport properties of the host crystal. We first investigate the polaron field in the non-interacting limit, and then include Coulomb and dispersive interactions leading to the formation of polaron pairs.

1. Rationale

In his pioneering work on the small polaron [1], Holstein suggested that the coupling to an asymmetric intramolecular vibration could split the vibronic potential energy into a double-well type "...*quite similar to that encountered in the theory of the inversion spectrum of the ammonia molecule*." Now that the modern theory of the ammonia spectrum is based on the pseudo-Jahn-Teller effect [2], Holstein's suggestion can be specified by considering the vibronic mixing by an odd-parity intermolecular mode of two nearly-degenerate opposite-parity electronic states at the small-polaron site. As the coupling strength (the Jahn-Teller energy) is increased, strong mixing arises which changes the vibronic potential-energy profile from single- to double- well. At this instant an elasto-electric dipole occurs due to the configurational transfiguring from centrosymmetric to non-centrosymmetric with the underlying parity non-conservation.

In what follows, we first introduce the concept of off-center small polarons in the localized limit leading to a charge-ordered state. At the beginning, we disregard the electrostatic interactions within the small-polaron field. Our investigation at this stage is carried out by the adiabatic approximation which separates the electronic and vibrational aspects of the overall problem [1]. As a result of interwell tunneling above the mode-softening point the total dipole vanishes, the tunneling tending to restore *on the average* the broken inversion symmetry at the polaron site. We define the vibronic ground-state tunneling splitting for the extrema of large and small polarons. At the small polaron limit, our result is similar to Holstein's formula obtained by perturbation theory, but his polaron reduction factor appears squared.

Next, we introduce the electrostatic interaction in a step-by-step approach. To define the quantities that effect the coupling of a quantized field of off-center small polarons to an external electric field, we place an extra charge at an arbitrary site in the host crystal. Expanding the interaction energy into a multipole series, we extract monopole -monopole intraband repulsion terms, as well as monopole-dipole interband attraction terms. The elasto-electric character of the latter terms is clearly seen: they rise from nil at the mode-softening point to gradually saturate finite-valued at the small-polaron extreme. Accounting for the dipole-smearing effect, due to the interwell tunneling, we also derive a vibronic polarizability for the localized off-center small polaron and show that its interaction with the extra charge is of the monopole - induced dipole type. At this point we estimate the contribution of the off-center dipoles to the dielectric constant of the host crystal.

To shape our localized-limit approach we consider pair interactions within the quantized filed of off-center small polarons. As above, we show these interactions to compose of intraband monopole-monopole repulsion terms, as well as of interband dipole-dipole interaction terms, either attractive or repulsive, resulting from the vibronic mixing above the mode-softening point. Again, taking into account the dipolar smearing we show the pairing (binding) energy to be of the Van der Waals interatomic type in which the electronic beating frequency and the polarizability are substituted for by their respective vibronic counterparts. Expected optical properties are also touched briefly.

We finally relax the localization condition to consider the complete Hamiltonian which obtains by adding up the electron-hopping part. By carrying out the conversion to polarons through the appropriate ellimination of the phonon coordinates, the relevant small-polaron Hamiltonian results in which the off-center dipole-coupling terms enter explicitly. We consider a two-mode lattice in which a local dipolar mode promotes the off-center displacement, while another itinerant mode promotes the translational (hopping) motion. To secure itinerancy, we require that the related polaron bands, local and translational, overlap. Certainly, our localized-limit solutions would hold good in case of a translational non-adiabaticity when the hopping is slow with respect to the vibrational frequency. Finally, we discuss problems related to the transport properties of off-center small polarons.

Throughout this study, we have adopted Holstein's linear-chain approach to the lattice resolving the local vibration into symmetric (even parity) and asymmetric (odd parity) components. Whereas the latter induce an off-center instability, the former promote the translational motion. We believe this general picture resembles closely Holstein's original ideas.

## 2. Off-center vibronic polarons
### 2.1. Noninteracting limit

To begin with, we first disregard both the itinerancy and the interparticle interactions to introduce the *localized vibronic small polaron*. Off-center *small polarons* may be expected to occur at centrosymmetric configurations as polaron sites are disturbed by coupling to odd-parity vibrations. Throughout this paper we assume that the driving force of the off-center instability is the quantum-mechanical pseudo-Jahn-Teller effect (PJTE) [3,4]. Accordingly, let there be two closely-spaced (nearly degenerate) opposite-parity narrow electron (hole) bands, labelled $\alpha,\beta = 1,2$ which mix up vibronically through coupling to an odd-parity vibrational mode $q_{\alpha\beta}$ ($\alpha \neq \beta$). In the tight-binding limit both $\alpha$ and $\beta$ will be regarded as nearly single levels. We start by considering the following adiabatic Hamiltonian:

$$H_{AD} = \sum_{i,\alpha} E_{i\alpha} a_{i\alpha}^\dagger a_{i\alpha} + \sum_I [½ K_{i\alpha\beta} q_{i\alpha\beta}^2 + G_{i\alpha\beta} q_{i\alpha\beta}(a_{i\alpha}^\dagger a_{i\beta} + a_{i\beta}^+ a_{i\alpha})] \qquad (1)$$

Here the subscripts i run over the small-polaron sites, $E_{i\alpha}$ are the one-electron energies, $a_{i\alpha}^{\dagger}$ ($a_{i\alpha}$) are the fermion creation (annihilation) operators at site i in band $\alpha$. $K_{i\alpha\beta}$, $G_{i\alpha\beta}$ and $q_{i\alpha\beta}$ are the spring- and (electron-phonon) mixing- constants, and the co-ordinates of the mixing mode, respectively. In so far as we are concerned with low-temperature phenomena involving ground vibrational states, introducing field operators for the vibrational counterpart in (1) may not be imminent. In the adiabatic approximation the local electron (hole) motion at any site is presumed fast enough so that the ionic environment cannot follow suit and responds to the average electronic distribution only.

Using (1) we solve Schrödinger's equation $H_{AD}\psi = E_{AD}\psi$ by means of a superposition of one-particle states in either band:

$$\psi = \sum_{i\alpha} c_{i\alpha} a_{i\alpha}^{\dagger} |0\rangle \qquad (2)$$

$|0\rangle$ is the electronic vacuum state when both bands are empty (electrons) or filled (holes). A secular equation obtains for $E_{AD}$ whose roots are

$$E_{AD}^{\pm}(q_{i\alpha\beta}) = \tfrac{1}{2}\{K_{i\alpha\beta}q_{i\alpha\beta}^2 \pm \sqrt{[(2G_{i\alpha\beta}q_{i\alpha\beta})^2 + (E_{i\beta} - E_{i\grave{a}})^2]}\} \qquad (3)$$

while the linear-combination coefficients in (2) corresponding to $E_{AD}^{-}$ and $E_{AD}^{+}$ read

$$c_{i1}^{-} = 2^{-\frac{1}{2}}[cos(\phi_{i\alpha\beta}/2) + sin(\phi_{i\alpha\beta}/2)]$$

$$c_{i1}^{+} = 2^{-\frac{1}{2}}[cos(\phi_{i\alpha\beta}/2) - sin(\phi_{i\alpha\beta}/2)]$$

$$c_{i2}^{-} = -2^{-\frac{1}{2}}[cos(\phi_{i\alpha\beta}/2) - sin(\phi_{i\alpha\beta}/2)]$$

$$c_{i2}^{+} = 2^{-\frac{1}{2}}[cos(\phi_{i\alpha\beta}/2) + sin(\phi_{i\alpha\beta}/2)] \qquad (4)$$

with

$$\phi_{i\alpha\beta} = tan^{-1}(E_{i\alpha\beta}/2G_{i\alpha\beta}q_{i\alpha\beta}), \quad E_{i\alpha\beta} = |E_{i\beta} - E_{i\alpha}| \qquad (5)$$

Equations (4) and (5) will be utilised to compute expectation values in the electronic state (2).

In the adiabatic approximation, equation (3) describes the vibronic potential energy of the electronically-coupled oscillator at site i. $E_{AD}^{-}(q_{i\alpha\beta})$ is a double-well potential with two symmetric equilibrium positions at $\pm q_{i\alpha\beta}$:

$$q_{i\alpha\beta 0} = \sqrt{[(G_{i\alpha\beta}/K_{i\alpha\beta})^2 - (E_{i\alpha\beta}/2G_{i\alpha\beta})^2]} \qquad (6)$$

Consequently the lower-branch vibronic Hamiltonian at a polaron site reads:

$$H_{vib0} = -(\eta^2/2M_{i\alpha\beta})(d^2/dq_{i\alpha\beta}^2) + \tfrac{1}{2}\{K_{i\alpha\beta}q_{i\alpha\beta}^2 - \sqrt{[(2G_{i\alpha\beta}q_{i\alpha\beta})^2 + E_{i\alpha\beta}^2]}\} \qquad (7)$$

Solving Schrödinger's equation $H_{vib0}\chi(q_{i\alpha\beta}) = E_{vib}\chi(q_{i\alpha\beta})$ by means of the linear combination:

$$\chi(q_{i\alpha\beta}) = A\chi_g(q_{i\alpha\beta}-q_{i\alpha\beta 0}) + B\chi_g(q_{i\alpha\beta}+q_{i\alpha\beta 0}) \tag{8}$$

where

$$\chi_g(q_{i\alpha\beta}) = (K_{i\alpha\beta}/\eta\omega_{i\alpha\beta})^{1/4} exp[-(K_{i\alpha\beta}/2\eta\omega_{i\alpha\beta})q_{i\alpha\beta}^2] \tag{9}$$

is the ground-state harmonic oscillator wave function centered at $q = 0$, we arrive at

$$E_{vib}^{\pm} = (1 - \pm S)^{-1}(H_{RR} - \pm H_{RL}) \tag{10}$$

with

$$S = exp(-\xi_{i\alpha\beta 0}^2), \quad \xi_{i\alpha\beta 0}^2 = (2E_{JTi\alpha\beta}/\eta\omega_{i\alpha\beta})[1 - (E_{i\alpha\beta}/4E_{JTi\alpha\beta})^2] \tag{11}$$

S is the overlap integral of the two harmonic-oscillator wave-functions, $\xi_{i\alpha\beta} = \sqrt{(K_{i\alpha\beta}/\eta\omega_{i\alpha\beta})} \times q_{i\alpha\beta}$ being the dimensionless vibrational co-ordinate. $E_{JTi\alpha\beta} = G_{i\alpha\beta}^2/2K_{i\alpha\beta}$ is the Jahn-Teller energy.

The matrix elements in (10) may be evaluated approximately through estimating the hopping integrals by the steepest-descent method at $8E_{JTi\alpha\beta}\eta\omega_{i\alpha\beta} > E_{i\alpha\beta}^2$:

$$I_1 = \pi^{-1/2} \int_{-\infty}^{+\infty} \sqrt{[(8E_{JTi\alpha\beta}/\eta\omega_{i\alpha\beta})(\xi + \xi_{i\alpha\beta 0})^2 + (E_{i\alpha\beta}/\eta\omega_{i\alpha\beta})^2]} \, exp(-\xi^2) d\xi$$

$$\sim [(8E_{JTi\alpha\beta}/\eta\omega_{i\alpha\beta})\xi_{i\alpha\beta 0}^2 + (E_{i\alpha\beta}/\eta\omega_{i\alpha\beta})^2]^{1/2} = 4E_{JTi\alpha\beta}/\eta\omega_{i\alpha\beta}$$

$$I_2 = \pi^{-1/2} \int_{-\infty}^{+\infty} \sqrt{[(8E_{JTi\alpha\beta}/\eta\omega_{i\alpha\beta})\xi^2 + (E_{i\alpha\beta}/\eta\omega_{i\alpha\beta})^2]} \, exp(-\xi^2) d\xi \sim E_{i\alpha\beta}/\eta\omega_{i\alpha\beta} \tag{12}$$

yielding

$$H_{RR} = \frac{1}{2}\eta\omega_{i\alpha\beta}(1 + \xi_{i\alpha\beta 0}^2 - I_1) \sim \frac{1}{2}\eta\omega_{i\alpha\beta}(1 + \xi_{i\alpha\beta 0}^2 - 4E_{JTi\alpha\beta}/\eta\omega_{i\alpha\beta})$$

$$H_{RL} = \frac{1}{2}\eta\omega_{i\alpha\beta}(1 - \xi_{i\alpha\beta 0}^2 - I_2)S \sim \frac{1}{2}\eta\omega_{i\alpha\beta}(1 - \xi_{i\alpha\beta 0}^2 - E_{i\alpha\beta}/\eta\omega_{i\alpha\beta})S \tag{13}$$

Equation (10) gives the energies of the ground-state tunneling levels of the double-well oscillator by first-order perturbation theory at small $E_{i\alpha\beta}$. The energy gap between them (full tunneling splitting) is

$$\Delta E = 2[SH_{RR}-H_{RL}]/(1-S^2) = \eta\omega_{i\alpha\beta}(2\xi_{i\alpha\beta 0}^2 - I_1 + I_2)S/(1-S^2)$$

$$\sim E_{i\alpha\beta}(1-E_{i\alpha\beta}/4E_{JTi\alpha\beta})/2\sinh(\xi_{i\alpha\beta 0}^2) \tag{14}$$

while the mid-gap energy (polaron bound energy) is

$E_P = [H_{RR} - SH_{RL}]/(1 - S^2) = \frac{1}{2} \eta\omega_{i\alpha\beta} [(1 + \xi_{i\alpha\beta 0}^2 - I_1) - (1 - \xi_{i\alpha\beta 0}^2 - I_2) S^2]/(1 - S^2)$

$\sim \frac{1}{2} \eta\omega_{i\alpha\beta} [1 + \xi_{i\alpha\beta 0}^2 - 4E_{JTi\alpha\beta}/\eta\omega_{i\alpha\beta} - (1 - \xi_{i\alpha\beta 0}^2 - E_{i\alpha\beta}/\eta\omega_{i\alpha\beta})S^2] / (1-S^2)$ (15)

For very strong PJTE coupling ($4E_{JTi\alpha\beta}/E_{i\alpha\beta} \gg 1$) equations (14) and (15) transform to

$\Delta E \sim E_{i\alpha\beta} \, exp(-2E_{JTi\alpha\beta}/\eta\omega_{i\alpha\beta})$

$E_P \sim \frac{1}{2}\eta\omega_{i\alpha\beta} - E_{JTi\alpha\beta}$ (16)

which is referred to as *the small-polaron extreme*. At the opposite weak-coupling end ($4E_{JTi\alpha\beta}/E_{i\alpha\beta} \geq 1$), further considerations based on equations (14) and (15) yield

$\Delta E \sim E_{i\alpha\beta}$

$E_P \sim \frac{1}{2} [\eta\omega_{i\alpha\beta} - E_{i\alpha\beta}]$ (17)

which is referred to as *the large-polaron extreme*. Here we remind that the polaron radius being inversely proportional to its binding energy, $\rho_P \sim -E_P^{-1}$, it explains the adopted terminology.

Equation (14) gives the full-width of the flip-flop polaron band for well interchange within an isolated configurational pair. Its simplified form (16) at small gaps $E_{i\alpha\beta}$ is particularly informative: By way of the polaron-reduction factor $exp(-2E_{i\alpha\beta}/\eta\omega_{i\alpha\beta})$, it projects the energy gap $E_{i\alpha\beta}$ into a polaron band $\Delta E$. The latter is the result of the vibronic mixing of two narrow bands separated by $E_{i\alpha\beta}$. At this point it is impending to clarify just what "small $E_{i\alpha\beta}$" means within our first-order perturbation procedure. A reasonable criterion for the fitness of the harmonic-oscillator wavefunction can be based on the actual well-bottom curvature at $q_{i\alpha\beta 0}$: $K_{i\alpha\beta 0} = K_{i\alpha\beta} [1 - (E_{i\alpha\beta}/4E_{JTi\alpha\beta})^2]$. A nearly-perfect harmonicity would require $K_{i\alpha\beta 0} \approx K_{i\alpha\beta}$ which implies $4E_{JTi\alpha\beta} \gg E_{i\alpha\beta}$, the small-polaron condition. The small parameter therefore is $E_{i\alpha\beta}/4E_{JTi\alpha\beta}$.

### 2.2. Coulomb interactions of small off-center vibronic polarons
#### 2.2.1. Monopole--induced-dipole coupling

To see just how a quantized field of off-center vibronic polarons responds to an external electric field, we imagine an extra electron (hole) at site l. Its interaction energy with the field will be

$H' = \sum_{ij\alpha\beta} W_{ij\alpha\beta} a_{i\alpha}^+ a_{j\beta}$ (18)

where the coupling constant is:[5]

$$W_{ij\alpha\beta} = \int w_\alpha^*(\mathbf{r}-\mathbf{r}_i) \, [(e_l e/\kappa) / |\mathbf{r}-\mathbf{r}_l|] \, w_\beta(\mathbf{r}-\mathbf{r}_j) d\mathbf{r} \qquad (19)$$

Here $\kappa$ is an appropriate dielectric constant, $w_\alpha(...)$, etc. are Wannier's functions of the host lattice. To compute (19) we introduce $\mathbf{r} - \mathbf{r}_i = \mathbf{u}_i$, $\mathbf{r}_l - \mathbf{r}_i = \mathbf{R}_i$ and incorporate a multipole expansion of the form

$$|\mathbf{r} - \mathbf{r}_l|^{-1} = |\mathbf{u}_i - \mathbf{R}_i|^{-1} \sim R_i^{-1}(1 - \mathbf{u}_i.\mathbf{R}_I/R_i^2) \quad \text{at} \quad u_i << R_i.$$

Inserting we get, up to the dipole terms,

$$W_{ij\alpha\beta} = [(e_l e/\kappa R_i)\delta_{\alpha\beta} - (e_l \mathbf{R}_i/\kappa R_i^3).\mathbf{p}_{i\alpha\beta}]\delta_{ij} \equiv [eU_i\delta_{\alpha\beta} - \mathbf{p}_{i\alpha\beta}.\mathbf{F}_i]\delta_{ij} \qquad (20)$$

where $U_i$ and $\mathbf{F}_i$ are the potential and the field created by the extra charge, respectively. $\mathbf{p}_{i\alpha\beta}$ is an $\alpha-\beta$ mixing dipole:

$$\mathbf{p}_{i\alpha\beta} = \int w_\alpha^*(\mathbf{u}_i) \, e\mathbf{u}_I \, w_\beta(\mathbf{u}_i) d\mathbf{u}_i \qquad (21)$$

Due to symmetry considerations, $\mathbf{p}_{\alpha\alpha} = \mathbf{0}$: This is the first time the parity condition has been used in our calculations. From the definition of a Wannier function,

$$w_\alpha(\mathbf{r}-\mathbf{r}_l) = N^{-\frac{1}{2}} \sum_{\mathbf{k}} u_{\mathbf{k}\alpha}(\mathbf{r}) \, exp[-i\mathbf{k}.(\mathbf{r}_l-\mathbf{r})], \qquad (22)$$

we also see that $p_{\alpha\beta} = -p_{\beta\alpha}$, provided $u_{\mathbf{k}\alpha}(...)$ are all real. We see that the interaction of an extra charge $e_l$ with the off-center polaron field is composed of intraband monopole-monopole terms $W_{ii\alpha\alpha} = e_l e/\kappa R_i$ and interband monopole-dipole terms $W_{ii\alpha\beta} = -\mathbf{p}_{i\alpha\beta}.\mathbf{F}_i$. For charges of the same sign the former are repulsive, while the latter are attractive.

In accordance with the adiabatic approximation formulae, we shall next compute the expectation value of the interaction Hamiltonian (18) in electronic state (2) (lower branch):

$$H_{ii\psi}' = <\psi|H_{ii}'|\psi> = 2W_{ii\alpha\beta}c_{i\alpha}^-c_{i\beta}^- = \mathbf{p}_{i\alpha\beta}.\mathbf{F}_I \, cos(\phi_{i\alpha\beta}) = \mathbf{p}_{i\alpha\beta}(q_{i\alpha\beta}).\mathbf{F}_i \qquad (23)$$

making use of equations (4) and (5) and taking up the interband terms. Here

$$\mathbf{p}_{i\alpha\beta}(q_{i\alpha\beta}) \equiv \mathbf{p}_{i\alpha\beta} \, cos(\phi_{i\alpha\beta}) = [2G_{i\alpha\beta}q_{i\alpha\beta}/(4G_{i\alpha\beta}^2 q_{i\alpha\beta}^2 + E_{i\alpha\beta}^2)^{\frac{1}{2}}] \, \mathbf{p}_{i\alpha\beta} \qquad (24)$$

is the adiabatic off-center dipole at $q_{i\alpha\beta}$. Its elasto-electric character is clearly seen because of the dependence on the configurational coordinate. As $q_{i\alpha\beta}$ oscillates between $-q_{i\alpha\beta0}$ and $+q_{i\alpha\beta0}$ across the interwell barrier, $p_{i\alpha\beta}(q_{i\alpha\beta})$ does so between $-p_{i\alpha\beta}(q_{i\alpha\beta0})$ and $+p_{i\alpha\beta0}(q_{i\alpha\beta0})$, and its average vanishes. Here at equilibrium

$$\mathbf{p}_{i\alpha\beta}(q_{i\alpha\beta0}) = \mathbf{p}_{i\alpha\beta}\,[(4E_{JTi\alpha\beta}/E_{i\alpha\beta})^2 - 1]^{\frac{1}{2}} / (4E_{JTi\alpha\beta}/E_{i\alpha\beta}) \qquad (25)$$

This dipole is only meaningful when so is the off-center displacement, that is, at $4E_{JTi\alpha\beta} > E_{i\alpha\beta}$. It arises at the mode-softening point $(4E_{JTi\alpha\beta} = E_{i\alpha\beta})$ and saturates to $\mathbf{p}_{i\alpha\beta}$ at the small-polaron extreme $(4E_{JTi\alpha\beta} >> E_{i\alpha\beta})$. Actually $\mathbf{p}_{i\alpha\beta}(q_{i\alpha\beta})$ is the average of $\mathbf{p}_{i\alpha\beta}(q_{i\alpha\beta})$ in the harmonic-oscillator ground state $\chi_g(q_{i\alpha\beta}-q_{i\alpha\beta 0})$ around the well-bottom. We shall also take this latter average converting $H_{ii\psi}'$ into $H_{ii\psi\chi}'$ by substituting dipole (25) for dipole (24) in equation (23) for the interaction Hamiltonian.

To see just how the vibronic ground-state tunneling splitting changes under the field $\mathbf{F}_i$ of the extra charge, we consider the complete vibronic Hamiltonian $H_{vib} = H_{vib0} + H_{\psi\chi}'$, where $H_{vib0}$ is given by equation (7), while $H_{\psi\chi}'$ obtains from equation (23). We introduce a new basis composed of the symmetric and antisymmetric combinations of $\chi_g(q_{i\alpha\beta}-q_{i\alpha\beta 0})$ and $\chi_g(q_{i\alpha\beta}+q_{i\alpha\beta 0})$: in this basis the field-coupling terms alone are off-diagonal. We obtain

$$E_{vib}\pm(\mathbf{F}) = \tfrac{1}{2}\{H_{SS}/(1-S) + H_{AA}/(1-S) \pm [\,[H_{SS}/(1-S)-H_{AA}/(1-S)]^2 +$$

$$4H_{SA}'H_{AS}'/(1-S^2)]^{\tfrac{1}{2}}\} \tag{26}$$

where

$$H_{SS} = H_{RR} + H_{RL},\ H_{AA} = H - H_{RL},\ H_{SA}'= H_{AS}'= \mathbf{p}_{i\alpha\beta}(q_{i\alpha\beta 0}).\mathbf{F}_i \tag{27}$$

From (26) we easily get

$$\Delta E(\mathbf{F}) = \{\Delta E(0)^2 + [2\mathbf{p}_{i\alpha\beta}(q_{i\alpha\beta 0}).\mathbf{F}_i]^2 / (1-S^2)\}^{\tfrac{1}{2}}$$

$$E_P(\mathbf{F}) = E_P(0) \tag{28}$$

where $\Delta E(0)$ is given by equation (14). The second term under the square root which amounts to $2p_{i\alpha\beta}(q_{i\alpha\beta 0}).\mathbf{F}_i$ at $S^2 \ll 1$ is the monopole–induced-dipole coupling potential. Inasmuch as the midgap energies are the same whether the field is on or off, the difference

$$\Delta E(\mathbf{F}) - \Delta E(0) = \Delta E(0)\{[1 + (2\mathbf{p}_{i\alpha\beta}(q_{i\alpha\beta 0}).\mathbf{F}_i/\Delta E(0))^2]^{\tfrac{1}{2}} -1\} \tag{29}$$

between the lowest-energy split-off levels can be regarded as part of the monopole-induced-dipole binding energy. Introducing a *vibronic polarizability*

$$\alpha_{vib} = [p_{i\alpha\beta}(q_{i\alpha\beta 0})^2/\Delta E(0)]_{av} \tag{30}$$

where the average is over the field-dipole angle, we get at small coupling strengths $2\mathbf{p}_{i\alpha\beta}(q_{i\alpha\beta 0}).\mathbf{F}_I / \Delta E(0) \ll 1$:

$$U_b = \tfrac{1}{2}\alpha_{vib} F_i^2, \tag{31}$$

the binding energy. Equation (31) is exactly of the form of the coupling energy of a polarizable species with an external electric field to which it couples through an induced dipole of magnitude $\alpha_{vib}F_i$.

### 2.2.2. Vibronic versus electronic polarizability

The vibronic polarizability $\alpha_{vib}$ of equation (30) should be compared with the electronic polarizability $\alpha_{el}$ of a system of two levels separated by $E_{i\alpha\beta}$:

$$\alpha_{el} = p_{i\alpha\beta}^2/E_{i\alpha\beta} \qquad (32)$$

From (25), (30), and (32) we get

$$\alpha_{vib}/\alpha_{el} = [1 - (E_{i\alpha\beta}/4E_{JTi\alpha\beta})^2][E_{i\alpha\beta}/\Delta E(0)] \qquad (33)$$

At the small-polaron extreme, $4E_{JTi\alpha\beta} \gg E_{i\alpha\beta}$, the above ratio is as large as $exp(2E_{JTi\alpha\beta}/\eta\omega_{i\alpha\beta})$ because of equation (16). This demonstrates convincingly the feasibility of considering the vibronic effects under the small-polaron conditions. These advantages disappear at the mode-softening point, $4E_{JTi\alpha\beta}/E_{i\alpha\beta} = 1$.

### 2.2.3. Dielectric constant

For an instant, we consider the complete localized-polaron Hamiltonian as obtained by performing the polaron renormalization in the adiabatic approximation:

$$H_{loc} = \Sigma_{i\alpha}(E_{i\alpha} - E_{JTi\alpha\beta})a_{i\alpha}^\dagger a_{i\alpha} + \Sigma_{i\alpha\beta}[eU_i\delta_{\alpha\beta} - \mathbf{p}_{i\alpha\beta}(q_{i\alpha\beta 0}).\mathbf{F}_i]a_{i\alpha}^\dagger a_{i\beta} \qquad (34)$$

where use has been made of equations (16) ($\eta\omega_{i\alpha\beta} \ll E_{JTi\alpha\beta}$) and (20) for small polarons. We next embed two extra charges $e_l$ and $e_m$ into the medium and see how their Coulomb interaction energy is affected by their coupling to the polaron field. Apart from the self-energy terms

$$-(\Delta E)^{-1}\Sigma_{i\alpha\beta}e_n^2[\mathbf{R}_i.\mathbf{p}_{i\alpha\beta}(q_{i\alpha\beta 0})]^2/R_i^6 \quad (n=l,m),$$

the intercharge coupling amounts to

$$W_{lm} = -\Sigma_i e_l e_m [2|p_{i\alpha\beta}(q_{i\alpha\beta 0})|^2/\Delta E][\mathbf{r}_i.(\mathbf{r}_i-\mathbf{R}_{lm})/|\mathbf{r}_i||\mathbf{r}_i-\mathbf{R}_{lm}|]$$

$$= -e_l e_m[8\pi\alpha_{vib}/v_p R_{lm}]$$

where $R_{lm}$ is the intercharge separation, while $v_p$ is the volume attributed to a small polaron ($v_p \sim N_p^{-1/3}$).[5] Only the interband terms have been retained. Subtracting from the intercharge coupling energy *in vacuo*, we obtain the net effect of the medium:

$$U_{lm} - W_{lm} = e_l e_m/k_{vib}R_{lm}$$

where the vibronic dielectric constant is

$$k_{vib} = (1 - 8\pi\alpha_{vib}/v_p)^{-1} \qquad (35)$$

This dielectric constant may be subject to an experimental determination.

### 2.2.4. Dipole-dipole coupling

Another example to be considered is the pair interaction of off-center small polarons. We define an interaction Hamiltonian

$$H' = \tfrac{1}{2}\sum_{\alpha\beta\gamma\delta ijlm} W_{\alpha\beta\gamma\delta ijlm}\, a_{i\alpha}^{\dagger} a_{j\beta}^{\dagger} a_{l\gamma} a_{m\delta} \tag{36}$$

where the coupling constant is

$$W_{\alpha\beta\gamma\delta ijlm} = \int w_{\alpha}^{*}(\mathbf{r}-\mathbf{r}_i)\, w_{\beta}^{*}(\mathbf{r'}-\mathbf{r}_j)[(e^2/\kappa)/|\mathbf{r}-\mathbf{r'}^3|]\, w_{\gamma}(\mathbf{r'}-\mathbf{r}_l)\, w_{\delta}(\mathbf{r}-\mathbf{r}_m)d\mathbf{r}d\mathbf{r'} \tag{37}$$

Introducing $\mathbf{r} - \mathbf{r}_i = \mathbf{u}_i$, $\mathbf{r'} - \mathbf{r}_j = \mathbf{v}_j$, $\mathbf{r}_i - \mathbf{r}_j = \mathbf{R}_{ij}$, we expand $|\mathbf{r}-\mathbf{r'}|^{-1} = |\mathbf{u}_i-\mathbf{v}_j+\mathbf{R}_{ij}|^{-1}$ into a multipole series. Considering co-linear displacements $u_i$ and $v_j$ that are both small ($u,v \ll R$) and correlated ($\mathbf{u}.\mathbf{R} = \mathbf{v}.\mathbf{R}$), we have

$$|\mathbf{r}-\mathbf{r'}|^{-1} = R_{ij}^{-1}\,[1+(u_i-v_j)/R_{ij}^2]^{-\tfrac{1}{2}} \sim R_{ij}^{-1}\,[1-(u_i-v_j)^2/2R_{ij}^2]$$

We also set $l = j$, $m = i$ (two sites) to obtain

$$W_{\alpha\beta\gamma\delta ijji} = (e^2/\kappa R_{ij})\delta_{\alpha\delta}\delta_{\beta\gamma} - (e^2/2\kappa R_{ij}^3)[<w_{\alpha}(u_i)|u_i|^2 w_{\delta}(u_i)>\delta_{\beta\gamma} +$$

$$<w_{\beta}(v_j)|v_j|^2 w_{\gamma}(v_j)>\delta_{\alpha\delta} - 2<w_{\alpha}(u_i)|{}^3u_i{}^3|w_{\delta}(u_i)><w_{\beta}(v_j)|v_j{}^3|w_{\gamma}(v_j)>] \tag{38}$$

From (38) and (21) we get a coupling constant of two-band species

$$W_{\alpha\beta\alpha\beta ijji} = (e^2/\kappa R_{ij})\delta_{\alpha\beta}\delta_{\beta\alpha} + (\mathbf{p}_{i\alpha\beta}.\mathbf{p}_{i\beta\alpha}/\kappa R_{ij}^3)(1 - \delta_{\alpha\beta}) -... \tag{39}$$

It is composed of an intraband ($\alpha=\beta$) monopole-monopole repulsion and an interband ($\alpha\neq\beta$) dipole-dipole term which is attractive because of $p_{i\alpha\beta} = -p_{i\beta\alpha}$. Other higher-order (quadrupole) intraband terms under the dots in (39) will not be taken into account presently. Interband terms like $W_{\alpha\beta\beta\alpha ijji}$ are also conceivable but they induce monopole-monopole repulsion mainly. Finally, combinations like $W_{\alpha\alpha\beta\beta ijji}$ induce monopole-monopole repulsion. For the purpose of seeking pairing mechanisms, therefore, we shall focus our attention onto equation (39). At $\alpha\neq\beta$ the corresponding interaction Hamiltonian reads

$$H' = \tfrac{1}{2}\sum_{\alpha\beta ij} [(\mathbf{p}_{i\alpha\beta}.\mathbf{p}_{j\beta\alpha}/\kappa R_{ij}^3)a_{i\alpha}^{\dagger}a_{j\beta}^{\dagger}a_{j\alpha}a_{i\beta} + h.c.] \tag{40}$$

accounting for the dipole-dipole terms alone.

We next average (40) over a two-particle electronic state $\psi$, which is the product of two one-particle states, each related to one of the two coupled oscillators at i and j, respectively:

$$H'_{\psi} = \tfrac{1}{2}\sum_{ij}<\psi_{ij}|H'_{ij}|\psi_{ij}> \tag{41}$$

with $|\psi_{ij}> = |\psi_i>|\psi_j>$ where $|\psi_{i,j}>$ are each given by (2). Regrouping the terms with i and j and comparing with (24) we get

$$H'_\psi = \tfrac{1}{2}\Sigma_{ij} \langle \psi_{ij} | (a_{i\alpha}^\dagger a_{i\beta} a_{j\beta}^\dagger a_{j\alpha} + \text{h.c.}) \psi_{ij} \rangle (\mathbf{p}_{i\alpha\beta} \cdot \mathbf{p}_{j\beta\alpha}/\kappa R_{ij})$$

$$= \Sigma_{ij}\, \mathbf{p}_{i\alpha\beta}(q_{i\alpha\beta}) \cdot \mathbf{p}_{j\beta\alpha}(q_{j\alpha\beta})/\kappa R_{ij}^3 \qquad (42)$$

Here $p_{i\alpha\beta}(q_{i\alpha\beta})$ is the adiabatic off-center dipole at $q_{i\alpha\beta}$.

The foregoing considerations leading to equation (28) hold good as well. We again select two sites at i and j and define the Hamiltonian of the two coupled oscillators therein by means of

$$H_{vib}(q_{i\alpha\beta}, q_{j\alpha\beta}) = H_{vib0}(q_{i\alpha\beta}) + H_{vib0}(q_{j\alpha\beta}) + H'_{ij\psi\chi} \qquad (43)$$

The corresponding zeroth-order vibrational eigenstate obtains as a product of $\chi(q_{i\alpha\beta})$ states. We next form symmetric and antisymmetric combinations of the one-oscillator states as before and then make use of their products. The latter diagonalize $H_{vib0}$, while $H'_{ij\psi\chi}$ is off-diagonal. After some lengthy calculations we obtain the pertinent *tunneling splitting*:

$$\Delta E(i,j) = 2\{\Delta E(i,i)^2 + [2\mathbf{p}_{i\alpha\beta}(q_{i\alpha\beta 0}) \cdot \mathbf{p}_{j\beta\alpha}(q_{j\beta\alpha 0})/\kappa R_{ij}^3]^2\}^{1/2} \qquad (44)$$

in which the first term under the square root is the contribution in the absence of coupling, while the second one is the coupling term. Proceeding just like in 2.2.1, we expand the square root at small coupling strengths (e.g. large $R_{ij}$) to obtain

$$\tfrac{1}{2}[\Delta E(i,j) - 2\Delta E(i,i)] = \Delta E(i,i)\{[1 + [2\,\mathbf{p}_{i\alpha\beta}(q_{i\alpha\beta 0}) \cdot \mathbf{p}_{j\beta\alpha}(q_{j\beta\alpha 0})/\kappa R_{ij}^3]/\Delta E(i,i)]^2]^{1/2} - 1\}$$

$$\approx \tfrac{1}{2}\Delta E(i,i)[2\mathring{a}_{vib}/\kappa R_{ij}^3]^2 = 4U_b$$

with

$$U_b = \tfrac{1}{2}\Delta E(i,i)[\mathring{a}_{vib}/\kappa R_{ij}^3]^2 \qquad (45)$$

standing for the vibronic Van der Waals binding energy.

### 2.2.5. Optical properties

Given an off-center vibronic polaron in ground state on the lower adiabatic sheet $E_{AD-}(q_{i\alpha\beta})$, it can absorb a light quantum and raise its energy to a quantized level pertaining to the upper sheet $E_{AD+}(q_{i\alpha\beta})$. There is a sole minimum on $E_{AD+}(q_{i\alpha\beta})$ at $q_{i\alpha\beta}=0$ corresponding to an inversion symmetry. One important conclusion, therefore, is that off-centerness is lost on optical excitation. Two peculiarities in the optical absorption spectrum may be expected to occur at photon energies of $4E_{JTi\alpha\beta}$ and $E_{JTi\alpha\beta} + \tfrac{1}{2}E_{i\alpha\beta}$, respectively. For an off-center species, more strongly localized in any of the side wells on $E_{AD-}(q_{i\alpha\beta})$ because of a higher barrier, the former energy will appear as a threshold in optical absorption, while the latter is not likely to occur at all. In any event, there will be an optical emission at photon energies about $E_{i\alpha\beta}$.

## 3. Itinerant off-center small polarons

We have so far considered the quantized field of off-centered species in the localized limit in which only a tunneling interchange between the off-center sites is allowed for. Adding itinerancy to the scheme would require the inclusion of the intersite hopping. We do this in writing down the complete Hamiltonian relevant to the present problem:

$$H = \sum_{(ij)\alpha\beta} t_{ij\alpha\beta}\, a_{i\alpha}^\dagger a_{j\beta} + \sum_{i\alpha} E_{i\alpha}\, a_{i\alpha}^\dagger a_{i\alpha} + \tfrac{1}{2}\sum_{ij\alpha\beta\gamma\delta} W_{ijji\alpha\beta\gamma\delta}\, a_{i\alpha}^\dagger a_{j\beta}^\dagger a_{j\gamma} a_{i\delta} -$$

$$\sum_i (\eta^2/2M_{i\alpha\beta})(\partial^2/\partial q_{i\alpha\beta}^2) + \tfrac{1}{2}\sum_i K_{i\alpha\beta} q_{i\alpha\beta}^2 + \sum_i G_{i\alpha\beta} q_{i\alpha\beta}(a_{i\alpha}^\dagger a_{i\beta} + \text{h.c.}) \qquad (46)$$

Here $t_{ij\alpha\beta}$ are the electron hopping integrals between neighboring sites (i,j), $E_{i\alpha}$ are the one-electron energies, $W_{ijji\alpha\beta\gamma\delta}$ are the pair coupling constants, and Greek superscripts label the two degenerate electron (hole) bands: $\alpha,\beta,\gamma,\delta = 1,2$ (*g,u*). The lattice quantities are regarded as c-numbers: $M_{i\alpha\beta}$, $K_{i\alpha\beta}$, and $q_{i\alpha\beta}$ are the masses, spring constants, and coordinates of the local oscillators at site i, while $G_{i\alpha\beta}$ are electron-lattice coupling constants. There are two such oscillators: one ($q_{i\alpha\alpha}$) securing translational itinerancy for $\alpha=\beta$ (even parity) and another one ($q_{i\alpha\beta}$) securing vibronic interband mixing for $\alpha\neq\beta$ (odd parity). The electron hopping integrals are also of two kinds: $t_{ij\alpha\beta}$ for $\alpha=\beta$ brings about the resonant intraband electron transfer (electronic bandwidths), while $t_{ij\alpha\beta}$ at $\alpha\neq\beta$ induces interband transitions (interband charge transfer). The energy reference level is at midpoint between the two electron (hole) bands.

Each electron (hole) band is assumed narrow. If both bands are sufficiently narrow so as to be inferior by width to the phonon energy coupled to the translational motion, $t_{ij\alpha\beta} < \eta\omega_{i\alpha\beta}$, then our *adiabatic-approximation*-based localized solutions of Section 2 would certainly apply in a first-order approach to the itinerant problem, provided the pair-coupling terms also posed a small perturbation. The above inequality defines the so-called *anti-adiabatic approximation* to the itinerancy [6]. Now performing the conversion to small polarons by following the prescriptions of Section 2, the complete Hamiltonian (46) transforms into

$$H = \sum_{(ij)\alpha\beta} t_{ij\alpha\beta}\, a_{i\alpha}^\dagger a_{j\beta} + \sum_{i\alpha} E_{i\alpha}\, a_{i\alpha}^\dagger a_{i\alpha} + \tfrac{1}{2}\sum_{ij\alpha\beta\gamma\delta} W_{ijji\alpha\beta\gamma\delta}\, a_{i\alpha}^\dagger a_{j\beta}^\dagger a_{j\gamma} a_{i\delta} \qquad (47)$$

where the renormalized quantities *in italics* are given by:

*t*$_{ij\alpha\beta}$ = t$_{ij\alpha\beta}$ *exp*(-2E$_{JTi\alpha\beta}$/η$\omega_{i\alpha\beta}$)

*t*$_{ii\alpha\beta}$ = t$_{\alpha\beta}$ *exp*(-2E$_{JTi\alpha\beta}$/η$\omega_{i\alpha\beta}$)(small polarons), = ½ΔE(0)(general)

*E*$_{i\alpha}$ = E$_{i\alpha}$ - E$_{iP}$

*W*$_{ijji\alpha\beta\gamma\delta}$ = U$_{ij\alpha\beta}$δ$_{\alpha\beta}$ + [2**p**$_{i\alpha\beta}$ (q$_{i\alpha\beta 0}$)·**p**$_{j\beta\alpha}$(q$_{j\alpha\beta 0}$)/κR$_{ij}^3$](1-δ$_{\alpha\beta}$),   (48)

etc. Here $t_{\alpha\beta} = \tfrac{1}{2}E_{\alpha\beta}$, while $U_{ij\alpha\beta}$ is the electron-electron correlation energy. Equations (48) give the renormalized polaron quantities. While *t*$_{ij\alpha\beta}$ is the itinerant hopping integral, *t*$_{ii\alpha\beta}$ is the one for reorientational well-interchange of the off-center polaron. $E_{iP}$ is the polaron binding energy, as obtained from equation (15). To secure the itinerancy, the two polaron bands, translational and reorientational, should overlap which is met if the respective Jahn-Teller energies are the same: $E_{JTi\alpha\beta} = E_{JTi\alpha\alpha} = E_{JT}$.

In a rigorous treatment, going from localized to itinerant limit would also require a change viz. by expanding the field operators from Wannier functions to Bloch functions: This would ultimately transfigure the meaning of the fermion creation (annihilation) operators $a^{\dagger}$ ($a$), which are the amplitudes of those expansions. Nevertheless, the former expansion can possibly be retained applying to both localization and itinerancy because translational hopping may be secured by the (small) overlap of Wannier's functions at neighboring sites along the lattice chain. We, therefore, assume once again that our localized solution of Section 2 is also a reasonable low-order approach to itinerancy in the anti-adiabatic approximation.

We finally present the relevant formulae for calculating the pertinent polaron masses for translational and reorientational motions (a tight binding approach):

$$m_{ij\alpha\beta} = \eta^2 / 2t_{ij\alpha\beta} d_{ij\alpha\beta}^2 = (\eta^2 / 2d_{ij\alpha\beta}^2 t_{ij\alpha\beta}) \exp(2E_{JTi\alpha\beta}/\eta\omega_{i\alpha\beta}) \qquad (49)$$

$$m_{ii\alpha\beta} = \eta^2 / 2t_{ii\alpha\beta} d_{ii\alpha\beta}^2 = (\eta^2 / 2d_{ii\alpha\beta}^2 t_{i\alpha\beta}) \exp(2E_{JTi\alpha\beta}/\eta\omega_{i\alpha\beta}), \qquad (50)$$

respectively (small polarons). Here $d_{ij\alpha\beta}$ are the corresponding hopping distances. Both masses are largely increased by the reciprocal polaron reduction factors. $\Delta E(0)$ is the zero-field tunneling splitting, as defined by equation (14). In an antiadiabatic approximation to itinerancy $m_{ij\alpha\beta} > m_{ii\alpha\beta}$, since the local adiabatic approximation requires $t_{\alpha\beta} > \eta\omega_{i\alpha\beta} > t_{ij\alpha\beta}$.

### 3.1. Conduction by off-center small polarons

Apart from carrying an inherent electric dipole at high reorientational barriers or an inherent vibronic polarizability at low barriers, the itinerant off-center polarons are charge carriers which transport electric currents across the crystal. In this respect, the species may be expected to exhibit nearly all the intrinsic features of small-polaron conduction [7], including pairing so as to form bipolarons [6]. However, the pairing mechanism is particularly effective in the off-centered case, as described by the dipole-dipole or *vibronic Van der Waals coupling* of 2.2.4. Unless the itinerant bipolaron mass is too large which would rather lead to a localized charge-ordered state, mobile bipolarons may Bose condense at some critical temperature to bring about an electronic superfluidity and, thereby, superconductivity to the host crystal. The presumed occurrence of superconductivity within a gas of small off-center polarons has been suggested as the possible mechanism in high-$T_c$ materials [8-10].

### 4. Discussion

We have virtually explored Holstein's idea by considering the electron coupling to two vibrational modes along a linear molecular chain. One of these (odd parity) brings about an off-center configurational distortion to the resulting small polaron, while the other (even parity) induces the translational motion along the chain of that polaron. Both off-centerness and itinerancy are regarded as vibronic effects resulting from interband and intraband mixing, respectively, involving two electron (hole) bands of different parities. In other words, while the off-center polaron occurs due to a *pseudo-Jahn-Teller effect*, its translational motion is described as a *dynamic Jahn-Teller effect*. The bare electron (hole) band, appearing as a resonance crossover splitting in the latter case, is assumed to be sufficiently narrow so as to justify applying the anti-adiabatic approach to the itinerancy. At the same time, the interband energy gap, entering as a crossover splitting in the former, is assumed sufficiently large to secure the validity of the adiabatic approximation to describing the intrasite configurational structure. Under these conditions the off-center structure formed at a given site may be expected to remain stable as the polaron hops to a neighboring site: In this way an off-center instability associated with charge carriers will travel along the crystalline chain. Alternative Green function approaches to similar dipolar arrangements in ferroelectric and antiferroelectric crystals have led, perhaps in a more stringent manner, to virtually the same frequency renormalization as the one obtained by the more transparent adiabatic-approximation-based localized-limit approach [4]. One remaining problem is to see just to what extent the off-center itinerancy is obliged to the anti-adiabatic conjecture, that is, whether it can survive lifting the narrow-band restriction.

Each individual off-center instability is associated with an elasto-electric dipole or, else, with a vibronic polarizability when the underlying interwell barrier is sufficiently low. Indeed as the barrier height grows, the system will tend to localize in one of the wells because the orientational bandwidth will become narrower. Consequently, its response to an external electric field will more likely be the one of a permanent rather than of a smeared-off dipole. One way or the other, an itinerant off-center instability would imply a migrating dipole and once two such dipoles get together sufficiently closely they will either repel or attract each other, depending on their relative orientation. The latter case provides a highly efficient pairing mechanism at low barriers when the vibronic polarizability determines the pairing energy, the *vibronic Van der Waals binding*. Actually, this is a very simple pairing model and it is rather difficult to argue why it should not materialize in reality. If the binding energy exceeds largely the polaron bandwidth, the pairs turn into bipolarons which are sort of electronic dimers or quasi-molecules [6]. Here too some problems remain, for these bipolarons are only approximately bosonic because such are the products of the fermion operators [5]. The vibronic Van der Waals binding should occur irrespective of the same charge on the pairing partners and, regarding them as "neutral polarizable entities" (each one comprising two opposite electric charges), it also incorporates the electron-electron correlation energy, which would thus need not be accounted for separately. In this sense the dipolar pairing mechanism provides a specific realization of the negative-U models [11].

Bound off-center small polarons have earlier been introduced in the literature relative to the optical properties of oxides [12]. The present investigation can also be regarded as an extension of this concept so as to include itinerancy. Essential experimental verification should involve optical, dielectric, and transport measurements, such as those outlined in 2.2.5, 2.2.3, and 3.1, respectively.